\documentclass[a4paper,12pt]{article}
\usepackage{latexsym}
\usepackage{amsmath}
\usepackage{amssymb}
\usepackage{amscd}

\newlength{\minitwocolumn}
\makeatletter
\@addtoreset{equation}{section}
\makeatother

\title{\bf 
The massless XXZ chain \\
with a boundary
}

\author{Takeo Kojima}
\date{\it
Department Mathematics,
College of Science and Technology,\\
Nihon University, Chiyoda-ku, Tokyo
101-0062, Japan\\~\\
{\rm \today}
}

\begin{document}
\maketitle

\begin{abstract}
We study the XXZ chain with a boundary at massless
regime $-1<\Delta<1$.
We give the free fields realizations of the 
boundary vacuum state and it's dual.
Using these realizations,
we give the integral representations for
the correlation functions.
\end{abstract}

~\\

\section{Introduction}
The one-dimentional massless spin $\frac{1}{2}$ XXZ chain 
with a boundary is a system described by the Hamiltonian,
\begin{eqnarray}
{\cal H}_B=
-\frac{1}{2}
\sum_{n=1}^\infty (
\sigma_{n+1}^x \sigma_{n}^x+\sigma_{n+1}^y \sigma_{n}^y
+\Delta \sigma_{n+1}^z \sigma_{n}^z
)+h \sigma_1^z,~~~-1<\Delta<1.\label{Hamiltonian}
\end{eqnarray}
Here the $\sigma_n^x, \sigma_n^y$ and $\sigma_n^z$
stand for the Pauli matrices acting on the $n$-th site
of the semi-infinite spin chain :
\begin{eqnarray}
\cdots \otimes {\mathbb{C}}^2 \otimes
\cdots \otimes {\mathbb{C}}^2 \otimes
{\mathbb{C}}^2 \otimes {\mathbb{C}}^2
.
\end{eqnarray}
In this paper we are interested in so-called massless regime :
$$-1<\Delta=-\cos\left(\frac{\pi}{\xi+1}
\right)<1,~~\xi {\rm~:~generic},$$
where the spectrum of the Hamiltonian 
(\ref{Hamiltonian}) is gapless.
In recent works 
\cite{JKM, JM, KY, K}
the various massless models ``without boundary''
were discussed, in the framework
of the free field approach.
In this paper we shall study
the massless model ``with a boundary'',
in the framework of the free field approach.

In the earlier work \cite{JKKKM}
the massive XXZ chain with a boundary 
was considered.
The diagonalization
of the Hamiltonian was obtained and
the integral representations of the
correlation functions were derived,
in the framework of the free field approach.
It's $U_q(\widehat{sl_n})$-generalization
was achieved in \cite{FK}.
In the work \cite{JKKMW},
Baxter's Corner Transfer Matrix Method
were extended to the boundary problem-
the XYZ model with a boundary.
The quantum Knizhnik-Zamolodchikov equation
with boundary reflection, which governs the
correlation functions, was derived.
An infinite product formula of the one-point
function was derived by solving the difference equations.
The Corner Trnsfer Matrix Method can be applied to
only massive models.
Fortunately, the massless XXZ spin is a limiting case of
the massive model- the XYZ model \cite{JKM}.
Therefore the correlation functions of the
massless XXZ spin with a boundary were
described by
the following systems of the difference equations,
which imply in particular the quantum Knizhnik-Zamolodchikov
equation with reflections.
\begin{eqnarray}
&&G(\beta_1,\cdots,\beta_{j+1},\beta_j,\cdots,\beta_{2N})
_{\cdots, \epsilon_{j+1}, \epsilon_j,\cdots}\nonumber
\\
&=&\sum_{\epsilon_j',\epsilon_{j+1}'=\pm}
R(\beta_j-\beta_{j+1})_{\epsilon_j \epsilon_{j+1}}
^{\epsilon_j', \epsilon_{j+1}'}
G(\beta_1,\cdots,\beta_{j},\beta_{j+1},\cdots,\beta_{2N})
_{\cdots, \epsilon_{j}', \epsilon_{j+1}', \cdots},
\label{Dif1}
\end{eqnarray}
and
\begin{eqnarray}
G(\beta_1, \cdots, \beta_{2N-1}, -\beta_{2N})
_{\epsilon_1, \cdots, \epsilon_{2N}}
=K(\beta_{2N})_{\epsilon_{2N}}^{\epsilon_{2N}}
G(\beta_1, \cdots, \beta_{2N-1}, \beta_{2N})
_{\epsilon_1, \cdots, \epsilon_{2N}},\nonumber\\
G(\beta_1+\pi i, \cdots, \beta_{2N-1}, \beta_{2N})
_{\epsilon_1, \cdots, \epsilon_{2N}}
=K(\beta_{1})_{-\epsilon_{1}}^{-\epsilon_{1}}
G(-\beta_1+\pi i, \beta_2, \cdots, \beta_{2N})
_{\epsilon_1, \cdots, \epsilon_{2N}}.\nonumber\\
\label{Dif2}
\end{eqnarray}
Here $G_N(\beta_1,\cdots,\beta_{2N})$
is a function with values in ${\mathbb{C}}^{\otimes 2N}$.
The matrix 
$R(\beta)$ denots the $R$-matrix defined in (\ref{def:R}).
The matrix $K(\beta)$
denotes the boundary $K$-matrix
defined in (\ref{def:K}).
In this paper we construct the integral reprersentations
of the $N$-point correlation functions,
which satisfy the above systems of difference
equations.

In this connection we should mention about
the works \cite{HSWY, FKQ}.
They constructed the
bosonizations of the boundary vacuum state
of the type-II vertex opertaors,
for the sine-Gordon model \cite{HSWY} and 
the $SU(2)$-invariant massive
thirring model \cite{FKQ},
by using Lukyanov's bosonizations of the vertex operators
with uv-cutoff \cite{Lu}.
Therefore their constructions started with bosons with
uv-cutoff, and
the form factors were derived after removing 
the cutoff parameter at final stage.
In this paper we prefer to work directly with operators
with the cutoff parameter removed \cite{JKM}, and construct 
the bosonizations of the boundary state associated with
the type-I vertex operators.

Now a few words about the organization of the paper.
In section 2 we formulate our problem.
In section 3 we construct the bosonizations
of the boundary vacuum state and it's dual state.
In section 4 we derived the integral representations
for the correlation functions.
In Appendix A we summarize the bosonizations
of the vertex operators \cite{JKM}.
In Appendix B we summariz the Multi-Gamma functions.

\section{Formulation}
The purpose of this section is to formulate
the problem.\\
Let us set the $R$-matrix as
\begin{eqnarray}
R(\beta)=r(\beta)
\left(\begin{array}{cccc}
1&&&\\
&b(\beta)
&
c(\beta)
&\\
&c(\beta)
&
b(\beta)
&\\
&&&1
\end{array}
\right), \label{def:R}
\end{eqnarray}
where we set the components as
\begin{eqnarray}
b(\beta)=-\frac{\displaystyle {\rm sh}\left(\frac{\beta}{\xi+1}\right)}
{\displaystyle {\rm sh}\left(\frac{\beta+\pi i}{\xi+1}\right)
}~~~,
c(\beta)=\frac{\displaystyle {\rm sh}\left(\frac{\pi i}{\xi+1}\right)}
{\displaystyle {\rm sh}\left(\frac{\beta+\pi i}{\xi+1}\right)
}.
\end{eqnarray}
Here we set
\begin{eqnarray}
r(\beta)=-
\frac{S_2(i\beta|2\pi,\pi(\xi+1))
S_2(-i\beta+\pi|2\pi,\pi(\xi+1))}
{S_2(-i\beta|2\pi,\pi(\xi+1))
S_2(i\beta+\pi|2\pi,\pi(\xi+1))
},
\end{eqnarray}
where $S_2(\beta|\omega_1 \omega_2)$ is the double sine 
function defined in Appendix B.
\\
Let $\{v_+,v_-\}$ denote the natural basis
of $V={\mathbb{C}}^2$.
When viewed as an operator on
$V \otimes V$, the matrix elements of $R(\beta)$
are defined by
\begin{eqnarray}
R(\beta)v_{k_1}\otimes v_{k_2}=
\sum_{j_1,j_2=\pm}
v_{j_1}\otimes v_{j_2}R(\beta)_{j_1 j_2}^{k_1 k_2}.
\end{eqnarray}
The $R$-matrix satisfies the Yang-Baxter equation :
\begin{eqnarray}
R_{12}(\beta_1-\beta_2)R_{13}(\beta_1-\beta_3)
R_{23}(\beta_2-\beta_3)=
R_{23}(\beta_2-\beta_3)R_{13}(\beta_1-\beta_3)
R_{12}(\beta_1-\beta_2).
\end{eqnarray}
The normalization factor $r_0(\beta)$ is so chosen that
the unitarity and crossing relations are
\begin{eqnarray}
R_{12}(\beta)R_{21}(-\beta)&=&id,\\
R(-\beta)_{j_1 j_2}^{k_1 k_2}&=&
R(\beta-\pi i)_{-k_2 j_1}^{-j_2 k_1}.
\end{eqnarray}
The commutation relation of the type-I vertex operator
$\Phi_j(\beta)$ is given by
\begin{eqnarray}
\Phi_{j_1}(\beta_1)\Phi_{j_2}(\beta_2)=
\sum_{k_1,k_2=\pm}
R(\beta_1-\beta_2)_{j_1,j_2}^{k_1,k_2}
\Phi_{k_2}(\beta_2)\Phi_{k_1}(\beta_1).\label{com:I}
\end{eqnarray}
The bulk scattering matrix is given by
\begin{eqnarray}
S(\beta)=s(\beta)\left(\begin{array}{cccc}
1&&&\\
&b'(\beta)&c'(\beta)\\
&c'(\beta)&b'(\beta)&\\
&&&1
\end{array}\right)
\end{eqnarray}
where we set the components as
\begin{eqnarray}
b'(\beta)=\frac{\displaystyle {\rm sh}
\left(\frac{\beta}{\xi}\right)}{
\displaystyle
{\rm sh}\left(\frac{i\pi-\beta}{\xi}\right)},~~
c'(\beta)=\frac{\displaystyle
{\rm sh}\left(\frac{\pi i}{\xi}\right)}{
\displaystyle
{\rm sh}\left(\frac{i\pi-\beta}{\xi}\right)}
\end{eqnarray}
Here we set
\begin{eqnarray}
s(\beta)=\frac{S_2(-i\beta|2\pi,\pi \xi)
S_2(\pi+i\beta|2\pi,\pi \xi)
}{
S_2(i\beta|2\pi,\pi \xi)
S_2(\pi-i\beta|2\pi,\pi \xi)
}.
\end{eqnarray}
The commutation relation of the type-II vertex operator
$\Psi_j(\beta)$
is given by
\begin{eqnarray}
\Psi_{j_1}(\beta_1)\Psi_{j_2}(\beta_2)=
\sum_{k_1,k_2=\pm}
S(\beta_1-\beta_2)_{j_1,j_2}^{k_1,k_2}
\Psi_{k_2}(\beta_2)\Psi_{k_1}(\beta_1).
\end{eqnarray}
The commutation relation of thpe-I and type-II vertex
operators is given by
\begin{eqnarray}
\Psi_{j_1}(\beta_1)
\Phi_{j_2}(\beta_2)=
j_1 j_2 {\rm tan}\left(\frac{\pi}{4}+i\frac{\beta_1-\beta_2}{2}
\right)\Phi_{j_2}(\beta_2)\Psi_{j_1}(\beta_1).
\end{eqnarray}
The free field realization of the vertex operators 
on the Fock space of boson was given in
\cite{JKM}.
We summarized the bosonization of the vertex operators
in Appendix A.

Let us set the boundary $K$-matrix by
\begin{eqnarray}
K(\beta)=k(\beta)\left(\begin{array}{cc}
1&0\\
0&
\frac{\displaystyle {\rm sh}\left(\frac{\mu+\beta}{\xi+1}\right)}
{\displaystyle
{\rm sh}\left(\frac{\mu-\beta}{\xi+1}\right)
}
\end{array}\right),\label{def:K}
\end{eqnarray}
where the normalization factor is given by
\begin{eqnarray}
k(\beta)=k_0(\beta)k_1(\beta),
\end{eqnarray}
where
\begin{eqnarray}
k_0(\beta)=\frac{S_2(-2i\beta+4\pi|4\pi,\pi(\xi+1))
S_2(2i\beta+3\pi|4\pi,\pi(\xi+1))
}{
S_2(2i\beta+4\pi|4\pi,\pi(\xi+1))
S_2(-2i\beta+3\pi|4\pi,\pi(\xi+1))
}
,\\
k_1(\beta)=\frac{
S_2(-i\beta+i\mu+\pi|2\pi,\pi(\xi+1))
S_2(i\beta+i\mu+2\pi|2\pi,\pi(\xi+1))
}{
S_2(i\beta+i\mu+\pi|2\pi,\pi(\xi+1))
S_2(-i\beta+i\mu+2\pi|2\pi,\pi(\xi+1))
}
.
\end{eqnarray}
The matrix elements $K(\beta)_{j}^k$ are
defined by
\begin{eqnarray}
K(\beta)v_k=\sum_{j=\pm}v_j K(\beta)_j^k.
\end{eqnarray}
The $R$-matrix and the $K$-matrix satisfy the Boundary
Yang-Baxter equation.
\begin{eqnarray}
K_2(\beta_2)R_{21}(\beta_1+\beta_2)K_1(\beta_1)
R_{12}(\beta_1-\beta_2)=
R_{21}(\beta_1-\beta_2)K_1(\beta_1)R_{12}(\beta_1+\beta_2)
K_2(\beta_2).
\end{eqnarray}
The normalization factor $k(\beta)$ is so chosen 
that
the boundary unitarity and the boundary crossing relations are
\begin{eqnarray}
K(\beta)K(-\beta)&=&id,\\
K\left(\beta+\frac{\pi i}{2}\right)_j^j&=&
\sum_{k=\pm}R(2\beta)_{k,-k}^{-j,j}
K\left(-\beta+\frac{\pi i}{2}\right)_k^k.
\end{eqnarray}
Let us set the renormalized transfer matrix
\begin{eqnarray}
{\cal T}_B(\beta)=g^{-1} \sum_{j=\pm}
\Phi_j^*(-\beta)K(\beta)_j^j
\Phi_j(\beta).
\label{Transfer}
\end{eqnarray}
Here we set the dual vertex operator as
\begin{eqnarray}
\Phi_j^*(\beta)=\Phi_{-j}(\beta+\pi i),~~(j=\pm).
\end{eqnarray}
The constant factor is given by 
\begin{eqnarray}
g=-\frac{1}{\pi i}e^{-(\gamma+{\rm ln}(\pi(\xi+1)))
\frac{\xi}{\xi+1}}
\sin\left(\frac{\pi}{\xi+1}
\right)
\left(\pi(\xi+1)\Gamma\left(\frac{1}{\xi+1}\right)\right)^2
\lim_{\beta \to 0}\frac{g(-\beta-\pi i)}{\beta},
\label{const:g}
\end{eqnarray}
where $g(\beta)$ is given in (\ref{def:g}).\\
The renormalized transfer matrix has the following relations.
\begin{eqnarray}
[{\cal T}_B(\beta_1), {\cal T}_B(\beta_2)]=0,~~
(\beta_1,\beta_2 \in {\mathbb{R}}),\\
{\cal T}_B(0)=id,~~{\cal T}_B(\beta){\cal T}_B(-\beta)=id,\\
{\cal T}_B(-\beta+\pi i)={\cal T}_B(\beta).
\end{eqnarray}
The Hamiltonian ${\cal H}_B$ (\ref{Hamiltonian}) and the
renormalized transfer matrix ${\cal T}_B(\beta)$ (\ref{Transfer})
are related through the formula, naively.
\begin{eqnarray}
\left(\frac{d}{d\beta}{\cal T}_B\right)(0)
\sim
{\cal H}_B.
\end{eqnarray}
Inspite of constructing the eigenstate of the Hamiltonian
(\ref{Hamiltonian}),
we solve the following eigenstate problem 
of the transfer matrix
${\cal T}_B(\beta)
$.
\begin{eqnarray}
{\cal T}_B(\beta)|B\rangle=|B\rangle,\\
\langle B|{\cal T}_B(\beta)=\langle B|.
\end{eqnarray}

\section{Boundary state}
In this section we invoke the bosonization method
to find the explicit formulae for the boundary state
$|B\rangle$, assuming uniqueness.
The boundary state is determined by the following
relation.
\begin{eqnarray}
{\cal T}_B(\beta)|B\rangle=|B\rangle.\label{relation:T}
\end{eqnarray}
Actiong the vertex opertaors $\Phi_j(-\beta)$
from the left, we have the equivalent relation.
\begin{eqnarray}
K(\beta)_j^j \Phi_j(\beta)|B\rangle
=\Phi_j(-\beta)|B\rangle,~~~(j=\pm).
\label{def:boundary}
\end{eqnarray}
Here we have used the duality relation.
\begin{eqnarray}
\Phi_j(\beta)\Phi_k^*(\beta)=g \times \delta_{j,k},~~~(j,k=\pm),
\end{eqnarray}
where $g$ is defined in (\ref{const:g}).\\
We make the ansatz that the boundary state has the
following form.
\begin{eqnarray}
|B\rangle=e^F|vac\rangle,
\end{eqnarray}
where
\begin{eqnarray}
F&=&\frac{1}{2}\int_0^\infty
\frac{A(t)}{[b(t),b(-t)]}b(-t)^2 dt +
\int_0^\infty
\frac{B(t)}{[b(t),b(-t)]}b(-t)dt.
\end{eqnarray}
When we set the coefficients as 
$A(t)=-1$ and
\begin{eqnarray}
B(t)=\frac{1}{t}\frac{\displaystyle
{\rm sh}\left(\frac{\pi t}{2}\right)
{\rm sh}\left((i\mu-\frac{\pi \xi}{2})t\right)
}{\displaystyle
{\rm sh}\left(\frac{\pi}{2}(\xi+1)t\right)
}-\frac{1}{t}
\frac{\displaystyle
{\rm sh}\left(\frac{\pi t}{2}\right)
{\rm sh}\left(\frac{\pi t}{4}\right)
{\rm ch}\left(\frac{\pi \xi t}{4}\right)
}{\displaystyle
{\rm sh}\left(\frac{\pi}{4}(\xi+1)t\right)
}
,
\end{eqnarray}
the boundary state $|B\rangle$ satisfy the characterizing
relation (\ref{def:boundary}).\\
Let us prove the relation (\ref{def:boundary}).
In what follows we use the abberiviations :
$\omega_1=2\pi, \omega_2=\pi(\xi+1)$, and
\begin{eqnarray}
U_+(\beta)=\exp\left(-\int_0^\infty
\frac{b(t)}{{\rm sh}\pi t}e^{i\beta t}dt\right),~~
U_-(\beta)=\exp\left(
\int_0^\infty
\frac{b(-t)}{{\rm sh}\pi t}e^{-i\beta t}dt
\right),\\
\bar{U}_+(\alpha)=\exp\left(
\int_0^\infty
\frac{b(t)}{{\rm sh}\frac{\pi}{2} t}e^{i\alpha t}dt
\right),~~
\bar{U}_-(\alpha)=\exp\left(
-\int_0^\infty
\frac{b(-t)}{{\rm sh}\frac{\pi}{2} t}e^{-i\alpha t}dt
\right).
\end{eqnarray}
In what follows we omit non-essential constant factors.\\
At first we explain the formulas of the form
\begin{eqnarray}
X(\beta_1)Y(\beta_2)=C_{XY}(\beta_1-\beta_2)
:X(\beta_1)X(\beta_2):,
\end{eqnarray}
where $X,Y=U_j$, and $C_{XY}(\beta)$ is a meromorphic
function on ${\mathbb{C}}$.
These formulae follow from the commutation relation
of the free bosons.
When we compute the contraction of the basic
operators,
we often encounter an integral
\begin{eqnarray}
\int_0^\infty
F(t)dt,
\end{eqnarray}
which is divergent at $t=0$.
Here we adopt the following prescription
for regularization :
it should be understood as the countour integral,
\begin{eqnarray}
\int_C F(t)\frac{{\rm log}(-t)}{2\pi i}dt, 
\end{eqnarray}
where the countour $C$ is given by
\\
~\\
~\\
\unitlength 0.1in
\begin{picture}(34.10,11.35)(17.90,-19.35)
%
\special{pn 8}%
\special{pa 5200 800}%
\special{pa 2190 800}%
\special{fp}%
\special{sh 1}%
\special{pa 2190 800}%
\special{pa 2257 820}%
\special{pa 2243 800}%
\special{pa 2257 780}%
\special{pa 2190 800}%
\special{fp}%
\special{pa 2190 1600}%
\special{pa 5190 1600}%
\special{fp}%
\special{sh 1}%
\special{pa 5190 1600}%
\special{pa 5123 1580}%
\special{pa 5137 1600}%
\special{pa 5123 1620}%
\special{pa 5190 1600}%
\special{fp}%
%
\special{pn 8}%
\special{pa 5190 1200}%
\special{pa 2590 1210}%
\special{fp}%
\put(25.9000,-12.1000){\makebox(0,0)[lb]{$0$}}%
%
\special{pn 8}%
\special{ar 2190 1210 400 400  1.5707963 4.7123890}%
\put(33.9000,-20.2000){\makebox(0,0){{\bf Contour} $C$}}%
\end{picture}%

~\\

The action of the basic operator on the boundary satate is
given by
\begin{eqnarray}
U_+(\beta)|B\rangle=Const.h(\beta)
U_-(-\beta)|B\rangle,
\end{eqnarray}
where 
\begin{eqnarray}
h(\beta)&=&\frac{\Gamma_2(-2i\beta+4\pi|2\omega_1,\omega_2)
\Gamma_2(-2i\beta+\pi(\xi+1)|2\omega_1,\omega_2)
}{
\Gamma_2(-2i\beta+3\pi|2\omega_1,\omega_2)
\Gamma_2(-2i\beta+\pi(\xi+1)+\pi|2\omega_1,\omega_2)
}\nonumber\\
&\times&
\frac{
\Gamma_2(-i\beta+i\mu+\pi|\omega_1,\omega_2)
\Gamma_2(-i\beta-i\mu+\pi(\xi+1)+\pi|\omega_1,\omega_2)
}{
\Gamma_2(-i\beta+i\mu+2\pi|\omega_1,\omega_2)
\Gamma_2(-i\beta-i\mu+\pi(\xi+1)|\omega_1,\omega_2)
}.
\end{eqnarray}
The function $h(\beta)$ satisfies
\begin{eqnarray}
K(\beta)_+^+=k(\beta)=k_0(\beta)k_1(\beta)=\frac{h(-\beta)}
{h(\beta)}.
\end{eqnarray}
We have
\begin{eqnarray}
h(-\beta)\Phi_+(\beta)|B\rangle=
Const.h(-\beta)h(\beta)U_-(\beta)U_-(-\beta)|B\rangle.
\end{eqnarray}
Now we have proved the ``$+$''-part of characterizing
relation of the boundary state (\ref{def:boundary}).
\begin{eqnarray}
K(\beta)_+^+\Phi_+(\beta)|B\rangle=
\Phi_+(-\beta)|B\rangle.
\end{eqnarray}
Next we shall prove the ``$-$''-part of (\ref{def:boundary}).\\
The commutation relation of the basic operator is
given by
\begin{eqnarray}
U_+(\beta_1)U_-(\beta_2)
=g(\beta_1-\beta_2)
U_-(\beta_2)U_+(\beta_1),
\end{eqnarray}
where we set
\begin{eqnarray}
g(\beta)=
e^{\gamma\frac{\xi}{2(\xi+1)}}
\frac{\Gamma_2(-i\beta+2\pi|\omega_1,\omega_2)
\Gamma_2(-i\beta+\pi(\xi+1)|\omega_1,\omega_2)
}{
\Gamma_2(-i\beta+\pi|\omega_1,\omega_2)
\Gamma_2(-i\beta+\pi(\xi+1)+\pi|\omega_1,\omega_2)
}.\label{def:g}
\end{eqnarray}
The action of the basic operator on the boundary state is
given by
\begin{eqnarray}
\bar{U}_+(\alpha)|B\rangle=
Const.I(\alpha)\bar{U}_-(-\alpha)|B\rangle,
\end{eqnarray}
where we set
\begin{eqnarray}
I(\alpha)=-\frac{2i\alpha}{\pi(\xi+1)}
\frac{\Gamma\left(\frac{-i\mu-i\alpha}{\pi(\xi+1)}
+1-\frac{1}{2(\xi+1)}\right)}
{\Gamma\left(
\frac{i\mu-i\alpha}{\pi(\xi+1)}+\frac{1}{2(\xi+1)}\right)}.
\end{eqnarray}
From direct calculation, we have
\begin{eqnarray}
&&h(\beta)^{-1}{\rm sh}\left(\frac{\mu+\beta}{\xi+1}\right)
\Phi_-(\beta)|B\rangle\nonumber\\
&=&Const.\times
{\rm sh}\left(\frac{\mu+\beta}{\xi+1}\right)
\int_{-\infty}^\infty d\alpha
\prod_{\epsilon_1,\epsilon_2=\pm}
\Gamma\left(\frac{i(\epsilon_1\alpha+\epsilon_2\beta)}{\pi(\xi+1)}
+\frac{1}{2(\xi+1)}\right)
\nonumber\\
&\times&
{\rm sh}\left(\frac{\alpha+\beta}{\xi+1}-\frac{\pi i}{2(\xi+1)}
\right)
\times
\alpha ~\frac{\Gamma\left(
\frac{-i\mu-i\alpha}{\pi(\xi+1)}+1-\frac{1}{2(\xi+1)}\right)}
{
\Gamma\left(
\frac{i\mu-i\alpha}{\pi(\xi+1)}+\frac{1}{2(\xi+1)}\right)
}
\nonumber\\
&\times&
U_-(\beta)U_-(-\beta)\bar{U}_-(\alpha)\bar{U}_-(-\alpha)
|B\rangle.
\end{eqnarray}
Note that 
the operator part $U_-(\beta)U_-(-\beta)\bar{U}_-(\alpha)
\bar{U}_-(-\alpha)
$ is invariant under the change of variables
$\alpha \leftrightarrow -\alpha,
\beta \leftrightarrow -\beta
$.\\
We get
\begin{eqnarray}
&&h(\beta)^{-1}{\rm sh}\left(\frac{\mu+\beta}{\xi+1}\right)
\Phi_-(\beta)|B\rangle-
h(-\beta)^{-1}{\rm sh}\left(\frac{\mu-\beta}{\xi+1}\right)
\Phi_-(-\beta)|B\rangle\nonumber\\
&=&Const.\times
{\rm sh}\left(\frac{2\beta}{\xi+1}\right)
\int_{-\infty}^\infty d\alpha
\prod_{\epsilon_1,\epsilon_2=\pm}
\Gamma\left(\frac{i(\epsilon_1\alpha+\epsilon_2\beta)}{\pi(\xi+1)}
+\frac{1}{2(\xi+1)}\right)\nonumber\\
&\times&\prod_{\epsilon=\pm}
\Gamma\left(\frac{i(-\mu+\epsilon\alpha)}{\pi(\xi+1)}
+1-\frac{1}{2(\xi+1)}\right)
\times\alpha \prod_{\epsilon=\pm}
{\rm sh}\left(\frac{\mu+\epsilon\alpha}{\xi+1}-\frac{\pi i}{2(\xi+1)}
\right)\nonumber\\
&\times&
U_-(\beta)U_-(-\beta)
\bar{U}_-(\alpha)\bar{U}_-(-\alpha)|B\rangle.
\end{eqnarray}
The integrand of (RHS) of the above equation
is anti-symmetric to a change of the variable 
$\alpha \leftrightarrow -\alpha$.
It means the left-hand side becomes zero
after taking integral.
Therefore we arrive at the following.
\begin{eqnarray}
h(-\beta){\rm sh}\left(\frac{\mu+\beta}{\xi+1}\right)
\Phi_-(\beta)|B\rangle
=
h(\beta){\rm sh}\left(\frac{\mu-\beta}{\xi+1}\right)
\Phi_-(-\beta)|B\rangle.
\end{eqnarray}
Now we have proved 
the ``$-$''-part of the characterizing
relation of the boundary vacuum
state
(\ref{def:boundary}).

The dual boundary state $\langle B|$ is determined by
the following relation.
\begin{eqnarray}
\langle B|{\cal T}_B(\beta)=
\langle B|.
\end{eqnarray}
Acting the dual veterx operator $\Phi_j^*(\beta)$
from right, we have the equivalent relation.
\begin{eqnarray}
\langle B|\Phi_j(-\beta+\pi i)K(\beta)_{-j}^{-j}
=\langle B |\Phi_j(\beta+\pi i),~~(j=\pm).
\label{def:dboundary}
\end{eqnarray}
We make the ansatz that the boundary state has the following form.
\begin{eqnarray}
\langle B|=\langle vac |e^G,
\end{eqnarray}
where
\begin{eqnarray}
G=\frac{1}{2}\int_0^\infty
\frac{C(t)}{[b(t),b(-t)]}b(t)^2 dt
+\int_0^\infty
\frac{D(t)}{[b(t),b(-t)]}b(t)dt.
\end{eqnarray}
When we set the coefficients  as
$C(t)=-e^{-2\pi t}$ and
\begin{eqnarray}
D(t)=
\frac{e^{-\pi t}}{t}
\frac{\displaystyle
{\rm sh}\left(\frac{\pi t}{2}\right)
{\rm sh}\left((i\mu-\frac{\pi \xi}{2}-\pi)t\right)
}{\displaystyle
{\rm sh}\left(\frac{\pi}{2}(\xi+1)t\right)
}+\frac{e^{-\pi t}}{t}
\frac{\displaystyle
{\rm sh}\left(\frac{\pi t}{2}\right)
{\rm sh}\left(\frac{\pi t}{4}\right)
{\rm ch}\left(\frac{\pi}{4}\xi t\right)
}{
\displaystyle
{\rm sh}\left(\frac{\pi}{4}(\xi+1)t\right)
},
\end{eqnarray}
the state $\langle B|$ satisfies the characterizing relation
(\ref{def:dboundary}).
It can be shown as the same manner as the case of the state
$|B\rangle$. Here we omit details.

\section{Correlation functions}
In this section we calculate the vacuum expectation values
of type-I vertex operators, and obtain them as integrals
of meromorphic functions involving Multi-Gamma functions.

We shall consider the $2N$-point functions defined by
\begin{eqnarray}
G_{\epsilon_1 \cdots \epsilon_{2N}}
(\beta_1, \cdots,\beta_{2N})=
\frac{\langle B|
\Phi_{\epsilon_1}(\beta_1)\cdots
\Phi_{\epsilon_{2N}}(\beta_{2N})
|B\rangle}{
\langle B|B\rangle
}.\label{correlation}
\end{eqnarray}
From the commutation relation of the vertex operators
(\ref{com:I}), the vacuum expectation values 
(\ref{correlation}) satisfy the $R$-matrix
symmetry (\ref{Dif1}).
From the reflection relations
(\ref{def:boundary}) and (\ref{def:dboundary}),
the vacuum expectation values 
(\ref{correlation}) satisfy the 
reflection conditions (\ref{Dif2}).

Specializing the spectral parameters,
they give multi-point correlation functions
of the local spin operators of the massless XXZ spin
with a boundary :
\begin{eqnarray}
G_{-\epsilon_1, \cdots, -\epsilon_{N},
\epsilon_N, \cdots ,\epsilon_1}
(\beta_1+\pi i, \cdots, \beta_{N}+\pi i,
\beta_N, \cdots, \beta_1
).
\end{eqnarray}
After specializing the spectral parameters as the above,
our integral representations 
of the correlation functions
(\ref{correlation}) can be compared with
the formulae for the correlation functions 
of the boundary XYZ spin, which can be derived by
mapping to the boundary SOS model \cite{MW}.
Y. Hara \cite{Hara} considered
the mapping method \cite{LP} of the boundary XYZ model,
and derived the explicit formulae of the one-point functions,
whose $N$-point generalization seems to be 
tedious but straightforward
\cite{LP2}.
Comparsion with two formulae is our future problem.

Now let us calculate the vacuum expectation value
(\ref{correlation}),
explicitly.\\
Fixing indexes
$\{\epsilon_1, \cdots, \epsilon_{2N}\}$,
let us denote by $A$ the index set
\begin{eqnarray}
A=\{a|\epsilon_a=-, 1\leq a \leq 2N\}.
\end{eqnarray}
In order to evaluate the expectation vbalue
(\ref{correlation}), we invoke the bosonization formulae
of the vertex operators and the boundary state.
By normal-ordering the product of vertex operators, we have
the following formula.

\begin{eqnarray}
&&G_{\epsilon_1 \cdots \epsilon_{2N}}
(\beta_1, \cdots,\beta_{2N})\nonumber\\
&=&
\prod_{1\leq b_1<b_2 \leq 2N}
\frac{\Gamma_2\left(i(\beta_{b_2}-\beta_{b_1})+2\pi
|\omega_1 \omega_2\right)
\Gamma_2\left(i(\beta_{b_2}-\beta_{b_1})+\pi(\xi+1)
|\omega_1 \omega_2\right)
}{
\Gamma_2\left(i(\beta_{b_2}-\beta_{b_1})+\pi
|\omega_1 \omega_2\right)
\Gamma_2\left(i(\beta_{b_2}-\beta_{b_1})+\pi(\xi+2)
|\omega_1 \omega_2\right)
}\nonumber\\
&\times&
\prod_{a \in A}\int_{-\infty}^\infty d\alpha_a
\prod_{a \in A}\Gamma\left(
\frac{i(\alpha_a-\beta_a)}{\pi(\xi+1)}+\frac{1}{2(\xi+1)}
\right)\Gamma\left(
\frac{i(\beta_a-\alpha_a)}{\pi(\xi+1)}+\frac{1}{2(\xi+1)}
\right)\nonumber
\\
&\times&
\prod_{a_1<a_2 \atop{a_1,a_2 \in A}}
\left\{
(\alpha_{a_2}-\alpha_{a_1})
\frac{\displaystyle
\Gamma\left(\frac{i(\alpha_{a_2}-\alpha_{a_1})}{\pi(\xi+1)}
+1-\frac{1}{\xi+1}\right)}{
\displaystyle
\Gamma\left(\frac{i(\alpha_{a_2}-\alpha_{a_1})}{\pi(\xi+1)}
+\frac{1}{\xi+1}\right)
}\right\}\nonumber\\
&\times&
\prod_{a<b \atop{a \in A, 1\leq b \leq 2N}}
\frac{\displaystyle
\Gamma\left(\frac{i(\beta_b-\alpha_a)}{\pi(\xi+1)}+
\frac{1}{2(\xi+1)}\right)}{
\displaystyle
\Gamma\left(\frac{i(\beta_b-\alpha_a)}{\pi(\xi+1)}+
1-\frac{1}{2(\xi+1)}\right)
}
\prod_{b<a \atop{a \in A, 1\leq b \leq 2N}}
\frac{\displaystyle
\Gamma\left(\frac{i(\alpha_a-\beta_b)}{\pi(\xi+1)}+
\frac{1}{2(\xi+1)}\right)}{
\displaystyle
\Gamma\left(\frac{i(\alpha_a-\beta_b)}{\pi(\xi+1)}+
1-\frac{1}{2(\xi+1)}\right)
}\nonumber
\\
&&~\times
J(\{\beta_b\}_{b=1}^{2N}|\{\alpha_a\}_{a \in A}).\nonumber\\
\end{eqnarray}
Here we set
\begin{eqnarray}
J(\{\beta_b\}_{b=1}^{2N}|\{\alpha_a\}_{a \in A})
=\frac{
\displaystyle
\langle B|\exp\left(
\int_0^\infty
X(t)b(-t)dt\right)
\exp\left(
\int_0^\infty
Y(t)b(t)dt
\right)
|B\rangle}{\langle B|B \rangle},
\end{eqnarray}
where
\begin{eqnarray}
X(t)&=&
\sum_{b=1}^{2N}
\frac{e^{-i\beta_b t}}{\displaystyle
{\rm sh}(\pi t)}
-
\sum_{a \in A}
\frac{e^{-i\alpha_a t}}{\displaystyle
{\rm sh}\left(\frac{\pi}{2} t\right)}.
\\
Y(t)&=&
-\sum_{b=1}^{2N}
\frac{e^{i\beta_b t}}{\displaystyle
{\rm sh}(\pi t)}
+
\sum_{a \in A}
\frac{e^{i\alpha_a t}}{\displaystyle
{\rm sh}\left(\frac{\pi}{2} t\right)}=-X^*(t).
\end{eqnarray}
Next we evaluate the quantity 
$J(\{\beta_b\}|\{\alpha_a\})$.
Using the completeness relation of the coherent states
\cite{JKKKM}, and
performing the integral calculations,
we have 
\begin{eqnarray}
&&J(\{\beta_b\}_{b=1}^{2N}|\{\alpha_a\}_{a \in A})\nonumber
\\
&=&\exp\left(
\int_0^\infty
\frac{1}{1-A(t)C(t)}
\frac{{\rm sh}\left(\frac{\pi t}{2}\right)
{\rm sh}\left(\pi t\right)
{\rm sh}\left(\frac{\pi \xi t}{2}\right)
}{
t
{\rm sh}\left(\frac{\pi}{2}(\xi+1)t\right)
}\right.
\nonumber\\
&\times&\left(
\frac{1}{2}C(t)X(t)^2+A(t)C(t)X(t)Y(t)+\frac{1}{2}
A(t)Y(t)^2
\right)\nonumber\\
&+&
\left.\int_0^\infty
\frac{1}{1-A(t)C(t)}
\left\{
(D(t)+C(t)B(t))X(t)+(B(t)+A(t)D(t))Y(t)
\right\}dt
\right).
\end{eqnarray}
Here $A(t), B(t), C(t)$ and $D(t)$ are the coefficient
functions
in the boundary state and it's dual.
It is evaluated as follows. 
\begin{eqnarray}
J(\{\beta_b\}|\{\alpha_a\})=
J_{\beta}(\{\beta_b\})
J_{\alpha}(\{\alpha_a\})
J_{\beta \alpha}(\{\beta_b\}|\{\alpha_a\}).
\end{eqnarray}
Here we set
\begin{eqnarray}
J_{\beta}(\{\beta_b\})&=&
\prod_{b=1}^{2N}
\frac{S_2(i\mu+i\beta_b+\pi|\omega_1 \omega_2)}{
S_2(i\mu+i\beta_b+2\pi|\omega_1 \omega_2)
}
\prod_{b=1}^{2N}
\sqrt{\frac{S_2(2i\beta_b+4\pi|2\omega_1 \omega_2)
S_2(-2i\beta_b+2\pi|2\omega_1 \omega_2)}{
S_2(2i\beta_b+3\pi|2\omega_1 \omega_2)
S_2(-2i\beta_b+\pi|2\omega_1 \omega_2)
}}
\nonumber\\
&\times&
\prod_{b=1}^{2N}
\sqrt{\frac{S_3(2i\beta_b+3\pi|\omega_1 \omega_1 \omega_2)
S_3(-2i\beta_b+\pi|\omega_1 \omega_1 \omega_2)}{
S_3(2i\beta_b+4\pi|\omega_1 \omega_1 \omega_2)
S_3(-2i\beta_b+2\pi|\omega_1 \omega_1 \omega_2)
}}\nonumber\\
&\times&
\prod_{1\leq b_1 < b_2 \leq 2N}
\frac{S_3(i(\beta_{b_1}+\beta_{b_2})+3\pi|
\omega_1\omega_1\omega_2))
S_3(-i(\beta_{b_1}+\beta_{b_2})+\pi|
\omega_1\omega_1\omega_2))
}{
S_3(i(\beta_{b_1}+\beta_{b_2})+4\pi|
\omega_1\omega_1\omega_2))
S_3(-i(\beta_{b_1}+\beta_{b_2})+2\pi|
\omega_1\omega_1\omega_2))
}\nonumber\\
&\times&
\prod_{1\leq b_1<b_2\leq 2N}\prod_{\epsilon=\pm}
\left\{
S_2(i\epsilon(\beta_{b_1}-\beta_{b_2}+\pi|\omega_1\omega_2))
\frac{S_3(i\epsilon(\beta_{b_1}-\beta_{b_2})+\pi|
\omega_1 \omega_1 \omega_2)}{
S_3(i\epsilon(\beta_{b_1}-\beta_{b_2})+4\pi|
\omega_1 \omega_1 \omega_2)
}
\right\},\nonumber\\
\end{eqnarray}
\begin{eqnarray}
J_{\beta \alpha}(\{\beta_b\}|\{\alpha_a\})
&=&
\prod_{b=1}^{2N}\prod_{a \in A}
\sqrt{\sin\left(\frac{i(\alpha_a+\beta_b)}{\xi+1}+\frac{\pi}
{2(\xi+1)}\right)}\nonumber\\
&\times&
\prod_{b=1}^{2N}
\prod_{a \in A}
\prod_{\epsilon=\pm}
\frac{1}{\sqrt{S_2(i\epsilon(\alpha_a+\beta_b)+\frac{\pi}{2}|
\omega_1 \omega_2)}}
\\
&\times&
\prod_{b=1}^{2N}
\prod_{a \in A}
\prod_{\epsilon=\pm}
\frac{1}{S_2(i\epsilon(\alpha_a-\beta_b)+\frac{\pi}{2}|
\omega_1 \omega_2)
\Gamma_1(i\epsilon(\alpha_a-\beta_b)+\frac{\pi}{2}|
\omega_2)
},\nonumber
\end{eqnarray}
and
\begin{eqnarray}
J_{\alpha}(\{\alpha_a\})&=&
\prod_{a \in A}
\frac{\sqrt{
\sin\left(
\frac{2i\alpha_a}{\xi+1}\right)
}}{\sin\left(\frac{i\alpha_a+i\mu}{\xi+1}+\frac{\pi}
{2(\xi+1)}\right)}\nonumber
\\
&\times&\prod_{a \in A}\prod_{\epsilon=\pm}
\frac{1}{\sqrt{S_2(2i\epsilon \alpha_a+\pi|\omega_1 \omega_2)}}
\frac{1}{\sqrt{S_2(2i\epsilon \alpha_a+\pi
(\xi+1)|\omega_1 \omega_2)}}\nonumber\\
&\times&
\prod_{a \in A}
\frac{1}{\sqrt{\Gamma_1(-2i\alpha+\pi \xi|\omega_2)}}
\frac{1}{\sqrt{S_2(2i\alpha_a+\pi(\xi+1)+\pi|\frac{\omega_1}{2}
\omega_2)}
}
\frac{1}{\sqrt{S_2(-2i\alpha_a+\pi\xi|\frac{\omega_1}{2}
\omega_2)}
}
\nonumber\\
&\times&
\prod_{a_1<a_2 \atop{a_1,a_2 \in A}}
\prod_{\epsilon=\pm}\sin\left(
\frac{i\epsilon(\alpha_{a_1}+\alpha_{a_2})}{\xi+1}
+\frac{\pi}{\xi+1}\right)^{-1}\nonumber\\
&\times&
\prod_{a_1<a_2 \atop{a_1,a_2 \in A}}
\frac{1}{S_2(i(\alpha_{a_1}+\alpha_{a_2})+\pi \xi|\frac{\omega_1}{2}
\omega_2)
S_2(-i(\alpha_{a_1}+\alpha_{a_2})+\pi \xi+2\pi|\frac{\omega_1}{2}
\omega_2)}
\nonumber\\
&\times&
\prod_{a_1<a_2 \atop{a_1,a_2 \in A}}
\prod_{\epsilon=\pm}
\frac{1}{S_2(i\epsilon(\alpha_{a_1}-\alpha_{a_2})+\pi(\xi+1)|
\frac{\omega_1}{2}\omega_2)}\nonumber\\
&\times&
\prod_{a_1<a_2 \atop{a_1,a_2 \in A}}
\prod_{\epsilon=\pm}
\frac{1}{\Gamma_1(i\epsilon(\alpha_{a_1}-\alpha_{a_2})+
\pi(\xi+1)|\omega_2)
\Gamma_1(i\epsilon(\alpha_{a_1}-\alpha_{a_2})+
\pi\xi|\omega_2)
}.
\end{eqnarray}
Here we omit an irrelevant constant.

In order to get the integral representations
of the form factors of the local spin operators,
we have to calculate
the vacuum expectation value of both type-I and type-II
vertex operators, and obtain them
as integrals of meromorphic functions of Multi-Gamma functions.
\begin{eqnarray}
F_{j_1 \cdots j_N}^{k_1 \cdots k_M}
(\gamma_1 \cdots \gamma_M |
\beta_1 \cdots \beta_{N})=
\frac{\langle B|
\Phi_{k_1}(\gamma_1)\cdots
\Phi_{k_{M}}(\gamma_{M})
\Psi_{j_1}(\beta_1)\cdots
\Psi_{j_N}(\beta_N)
|B\rangle}{
\langle B|B\rangle
}.
\end{eqnarray}
Caluculation of the vacuum expectation values
is tedious but straightforward.
We can perform it as the same manner as the correlation functions.

~\\
~\\
{\bf Acknowledgements.}~~
This work was partly supported by Grant-in-Aid for
Encouragements for Young Scientists ({\bf A})
from Japan Society for the Promotion of Science. (11740099)

\begin{appendix}
\section{Vertex Operators}
Here we summarize the bosonizations
of the vertex operators \cite{JKM}.\\
Let us set free bosons $b(t) (t \in {\mathbb{R}})$
which satisfy
\begin{eqnarray}
[b(t),b(t')]=\frac{\displaystyle
{\rm sh}\left(\frac{\pi t}{2}\right) 
{\rm sh}(\pi t) 
{\rm sh}\frac{\pi t \xi}{2} 
}{
\displaystyle
t {\rm sh}\frac{\pi t (\xi+1)}{2} }\delta(t+t').
\end{eqnarray}
Let us set $a(t)$ by
\begin{eqnarray}
b(t){\rm sh}\frac{\pi t (\xi+1)}{2}=
a(t){\rm sh}\frac{\pi t \xi}{2}.
\end{eqnarray}
Let us consider the Fock space ${\cal H}$ generated by
the vacuum $|vac\rangle$ which satisfies
\begin{eqnarray}
b(t)|vac \rangle=0 ~~~{\rm if}~~t>0.
\end{eqnarray}
The bosonization of the type-I vertex operators
is given by
\begin{eqnarray}
\Phi_+(\beta)&=&U(\beta),\\
\Phi_-(\beta)&=&\int_{C_I}d\alpha
:U(\beta)\bar{U}(\alpha):\nonumber\\
&\times&
\Gamma\left(\frac{i(\alpha-\beta)}{\pi(\xi+1)}+
\frac{1}{2(\xi+1)}\right)
\Gamma\left(-\frac{i(\alpha-\beta)}{\pi(\xi+1)}+
\frac{1}{2(\xi+1)}\right),
\end{eqnarray}
where we have set
\begin{eqnarray}
U(\alpha)=:\exp\left(-
\int_{-\infty}^\infty \frac{b(t)}{{\rm sh} \pi t}
e^{i\alpha t}dt\right):,~~
\bar{U}(\alpha)=:\exp\left(
\int_{-\infty}^\infty \frac{b(t)}{{\rm sh} \frac{\pi}{2} t}
e^{i\alpha t}dt\right):.
\end{eqnarray}
The bosonization of the type-II vertex operators is
given by
\begin{eqnarray}
\Psi_+(\beta)&=&V(\beta),\\
\Psi_-(\beta)&=&\int_{C_{II}}d\alpha
:V(\beta)\bar{V}(\alpha):\nonumber\\
&\times&
\Gamma\left(\frac{i(\alpha-\beta)}{\pi \xi}-
\frac{1}{2 \xi}\right)
\Gamma\left(-\frac{i(\alpha-\beta)}{\pi \xi}+
-\frac{1}{2\xi}\right),
\end{eqnarray}
where we have set
\begin{eqnarray}
V(\alpha)=:exp\left(
\int_{-\infty}^\infty \frac{a(t)}{{\rm sh} \pi t}
e^{i\alpha t}dt\right):,~~
\bar{V}(\alpha)=:exp\left(
-\int_{-\infty}^\infty \frac{a(t)}{{\rm sh} \frac{\pi}{2} t}
e^{i\alpha t}dt\right):.
\end{eqnarray}
Here the integration contours are chosen as follows.
The contour $C_I$ is $(-\infty, \infty)$.
The poles
\begin{eqnarray}
\alpha-\beta=\frac{\pi i}{2}+n\pi(\xi+1)i,~~(n \in {\mathbb{N}})
\end{eqnarray}
of $\Gamma\left(\frac{i(\alpha-\beta)}{\pi(\xi+1)}+
\frac{1}{2(\xi+1)}\right)$ are above $C_I$
and
the poles
\begin{eqnarray}
\alpha-\beta=-\frac{\pi i}{2}-n\pi(\xi+1)i,~~(n \in {\mathbb{N}})
\end{eqnarray}
of $\Gamma\left(-\frac{i(\alpha-\beta)}{\pi(\xi+1)}+
\frac{1}{2(\xi+1)}\right)$ are below $C_I$.
The contour $C_{II}$ is $(-\infty, \infty)$
except that
the poles
\begin{eqnarray}
\alpha-\beta=-\frac{\pi i}{2}+n\pi \xi i,~~(n \in {\mathbb{N}})
\end{eqnarray}
of $\Gamma\left(\frac{i(\alpha-\beta)}{\pi \xi}-
\frac{1}{2 \xi}\right)$ are above $C_{II}$
and
the poles
\begin{eqnarray}
\alpha-\beta=\frac{\pi i}{2}-n\pi \xi i,~~(n \in {\mathbb{N}})
\end{eqnarray}
of $\Gamma\left(-\frac{i(\alpha-\beta)}{\pi \xi}-
\frac{1}{2 \xi}\right)$ are below $C_{II}$.

\section{Multi Gamma functions}

Here we summarize the multiple gamma and the multiple sine
functions.\\
Let us set the functions
$\Gamma_1(x|\omega), \Gamma_2(x|\omega_1, \omega_2)
$ and
$\Gamma_3(x|\omega_1, \omega_2, \omega_3)
$
by
\begin{eqnarray}
{\rm log}\Gamma_1(x|\omega)+\gamma B_{11}(x|\omega)&=&
\int_C\frac{dt}{2\pi i t}e^{-xt}
\frac{{\rm log}(-t)}{1-e^{-\omega t}},\\
{\rm log}\Gamma_2(x|\omega_1, \omega_2)
-\frac{\gamma}{2} B_{22}(x|\omega_1, \omega_2)&=&
\int_C\frac{dt}{2\pi i t}e^{-xt}
\frac{{\rm log}(-t)}{(1-e^{-\omega_1 t})
(1-e^{-\omega_2 t})},\\
{\rm log}\Gamma_3(x|\omega_1, \omega_2, \omega_3)
+\frac{\gamma}{3!} B_{33}(x|\omega_1, \omega_2, \omega_3)&=&
\int_C\frac{dt}{2\pi i t}e^{-xt}
\frac{{\rm log}(-t)}{(1-e^{-\omega_1 t})
(1-e^{-\omega_2 t})
(1-e^{-\omega_3 t})},\nonumber
\\
\end{eqnarray}
where
the functions $B_{jj}(x)$ are the multiple Bernoulli polynomials
defined by
\begin{eqnarray}
\frac{t^r e^{xt}}{
\prod_{j=1}^r (e^{\omega_j t}-1)}=
\sum_{n=0}^\infty
\frac{t^n}{n!}B_{r,n}(x|\omega_1 \cdots \omega_r),
\end{eqnarray}
more explicitly
\begin{eqnarray}
B_{11}(x|\omega)&=&\frac{x}{\omega}-\frac{1}{2},\\
B_{22}(x|\omega)&=&\frac{x^2}{\omega_1 \omega_2}
-\left(\frac{1}{\omega_1}+\frac{1}{\omega_2}\right)x
+\frac{1}{2}+\frac{1}{6}\left(\frac{\omega_1}{\omega_2}
+\frac{\omega_2}{\omega_1}\right).
\end{eqnarray}
Here $\gamma$ is Euler's constant,
$\gamma=\lim_{n\to \infty}
(1+\frac{1}{2}+\frac{1}{3}+\cdots+\frac{1}{n}-{\rm log}n)$.\\
Here the contor of integral is given by

~\\
~\\

\unitlength 0.1in
\begin{picture}(34.10,11.35)(17.90,-19.35)
%
\special{pn 8}%
\special{pa 5200 800}%
\special{pa 2190 800}%
\special{fp}%
\special{sh 1}%
\special{pa 2190 800}%
\special{pa 2257 820}%
\special{pa 2243 800}%
\special{pa 2257 780}%
\special{pa 2190 800}%
\special{fp}%
\special{pa 2190 1600}%
\special{pa 5190 1600}%
\special{fp}%
\special{sh 1}%
\special{pa 5190 1600}%
\special{pa 5123 1580}%
\special{pa 5137 1600}%
\special{pa 5123 1620}%
\special{pa 5190 1600}%
\special{fp}%
%
\special{pn 8}%
\special{pa 5190 1200}%
\special{pa 2590 1210}%
\special{fp}%
\put(25.9000,-12.1000){\makebox(0,0)[lb]{$0$}}%
%
\special{pn 8}%
\special{ar 2190 1210 400 400  1.5707963 4.7123890}%
\put(33.9000,-20.2000){\makebox(0,0){{\bf Contour} $C$}}%
\end{picture}%

~\\
~\\

Let us set
\begin{eqnarray}
S_1(x|\omega)&=&\frac{1}{\Gamma_1(\omega-x|\omega)
\Gamma_1(x|\omega)},\\
S_2(x|\omega_1,\omega_2)&=&\frac{
\Gamma_2(\omega_1+\omega_2-x|\omega_1,\omega_2)}{
\Gamma_2(x|\omega_1,\omega_2)},\\
S_3(x|\omega_1,\omega_2,\omega_3)&=&\frac{1}{
\Gamma_3(\omega_1+\omega_2+\omega_3-x|\omega_1,\omega_2,\omega_3)
\Gamma_3(x|\omega_1,\omega_2,\omega_3)}
\end{eqnarray}
We have
\begin{eqnarray}
\Gamma_1(x|\omega)=e^{(\frac{x}{\omega}-\frac{1}{2}){\rm log}
\omega}\frac{\Gamma(x/\omega)}{\sqrt{2\pi}},~
S_1(x|\omega)=2{\rm sin}(\pi x/\omega),
\end{eqnarray}
\begin{eqnarray}
\frac{\Gamma_2(x+\omega_1|\omega_1,\omega_2)}{
\Gamma_2(x|\omega_1,\omega_2)}=\frac{1}{\Gamma_1(x|\omega_2)},~
\frac{S_2(x+\omega_1|\omega_1,\omega_2)}{
S_2(x|\omega_1,\omega_2)}=\frac{1}{S_1(x|\omega_2)},~
\frac{\Gamma_1(x+\omega|\omega)}{\Gamma_1(x|\omega)}=x.
\end{eqnarray}

\begin{eqnarray}
\frac{
\Gamma_3(x+\omega_1|\omega_1,\omega_2, \omega_3)
}{\Gamma_3(x|\omega_1,\omega_2, \omega_3)}
=\frac{1}{\Gamma_2(x|\omega_2, \omega_3)},~
\frac{S_3(x+\omega_1|\omega_1,\omega_2,\omega_3)}{
S_3(x|\omega_1,\omega_2,\omega_3)}=\frac{1}{S_2(x|\omega_2,
\omega_3)}.
\end{eqnarray}

\begin{eqnarray}
{\rm log}S_2(x|\omega_1 \omega_2)
=\int_C \frac{{\rm sh}(x-\frac{\omega_1+\omega_2}{2})t}
{2{\rm sh}\frac{\omega_1 t}{2}
{\rm sh}\frac{\omega_2 t}{2}
}{\rm log}(-t)\frac{dt}{2\pi i t},~(0<{\rm Re}x<
\omega_1+\omega_2).
\end{eqnarray}

\begin{eqnarray}
S_2(x|\omega_1 \omega_2)=
\frac{2\pi}{\sqrt{\omega_1 \omega_2}}x +O(x^2),~~(x \to 0).
\end{eqnarray}

\begin{eqnarray}
S_2(x|\omega_1 \omega_2)
S_2(-x|\omega_1 \omega_2)=-4
{\rm sin}\frac{\pi x}{\omega_1}
{\rm sin}\frac{\pi x}{\omega_2}.
\end{eqnarray}
\end{appendix}
\end{document}